\documentclass[onecolumn,aps,superscriptaddress,prd,10pt,showpacs,preprintnumbers,nofootinbib,eqsecnum]{revtex4-2}
\usepackage{graphicx,slashed,bbold,subfigure,amsmath,amssymb,enumerate,color,xcolor}
\usepackage{multirow}
\usepackage{natbib,ulem} 
\usepackage[colorlinks=true,citecolor=blue,linkcolor=blue,urlcolor=blue]{hyperref}
\setcitestyle{square, numbers, comma, sort&compress}

\newcommand\Rsout{\bgroup\markoverwith{\textcolor{red}{\rule[0.5ex]{2pt}{0.4pt}}}\ULon}

\begin{document}
  
\title{Comparing strange and non-strange quark stars within resummed QCD at NLO} 
 \author{Tulio E. Restrepo}
  \email{trestre2@central.uh.edu}
  \affiliation{Department of Physics, University of Houston, Houston, TX 77204, USA}
 \affiliation{Instituto de F\'isica, Universidade Federal do Rio de Janeiro, Caixa Postal 68528, 21941-972, Rio de Janeiro, RJ, Brazil}

 \author{Jean-Lo\"{\i}c Kneur}
\email{jean-loic.kneur@umontpellier.fr}
\affiliation{Laboratoire Charles Coulomb (L2C), UMR 5221 CNRS-Universit\'{e} Montpellier, 34095 Montpellier, France}

 \author{Constan\c{c}a Provid\^{e}ncia}
  \email{cp@uc.pt}
  \affiliation{CFisUC, Department of Physics, University of Coimbra, 3004-516 Coimbra, Portugal}

  \author{Marcus Benghi Pinto}
 \email{ marcus.benghi@ufsc.br}
 \affiliation{Departamento de F\'isica, Universidade Federal de Santa Catarina, 88040-900 Florian\'opolis, SC, Brazil}

\begin{abstract}
We employ the {\it renormalization group optimized perturbation theory} (RGOPT)
resummation method to evaluate the equation of state (EoS) for  strange
($N_f=2+1$) and non-strange ($N_f=2$) cold quark matter at
next-to-leading order (NLO). This allows us to obtain the mass-radius relation
for  pure quark stars
and compare the results  with the predictions from perturbative QCD (pQCD) at NNLO. Choosing the renormalization scale to generate maximum star  masses of order $M=2 - 2.6 M_\odot$, we show that the RGOPT can produce  mass-radius curves compatible with the masses and radii of some recently observed pulsars,  regardless of their strangeness content. The scale values required to produce the desired maximum masses are higher in the strange scenario since the EoS  is softer in this case. The possible reasons for such behavior are discussed. Our results also show that, as expected, the RGOPT predictions for the relevant observables are  less sensitive to scale variations than those furnished by pQCD.   
\end{abstract}
\maketitle
\section{Introduction}

The recent observational data of nearly two solar mass pulsars \cite{Demorest:2010bx,Fonseca:2016tux,Antoniadis:2013pzd,Romani:2021xmb} suggests that quark matter (QM) may be present inside compact stellar objects (CSO). The theoretical description of those objects requires the knowledge of the quantum chromodynamics (QCD) equation of state (EoS) which can be determined by means of effective models \cite{Oertel:2016bki}, Bayesian inference  \cite{Lim:2019som,Traversi:2020aaa,Zhu:2022ibs,Malik:2022zol,Gorda:2023usm,Malik:2023mnx,Takatsy:2023xzf,Char:2023fue,Albino:2024ymc}, as well as  perturbative approximations to QCD, among other possibilities. The vast majority of those applications take the Bodmer–Witten hypothesis \cite{Bodmer:1971we,Teraza1979,Witten:1984rs} into account so that both, quark and neutron stars, are generally described in terms of strange quark matter (SQM), see for instance \cite{Glendenning:2000}. However, although mostly ignored in  the  literature, the intriguing possibility that quark matter may not be strange, so  that two-flavor quark stars with a larger maximum mass, can  exist  has also been advocated \cite{Holdom:2017gdc}.  Inspired by  this argument, the authors of Ref. \cite {Zhao:2019xqy} have used an alternative self-consistent mean field approximation  to study the EoS of cold dense matter within the framework of the two-flavor Nambu--Jona-Lasinio model (NJL) \cite{,Nambu:1961tp,Nambu:1961fr} (in recent years, more works exploring the properties of non-strange quark matter (NSQM) have come out \cite{Wang:2019jze,Zhang:2019mqb,Ren:2020tll,Wang:2020wzs,Xu:2021alh,Zhang:2020jmb,Yuan:2022dxb,Cao:2020zxi,Xia:2022tvx,Restrepo:2022wqn}). 
The results indicate that this approach can generate an EoS which allows for the existence of two-flavor quark stars whose maximum masses can be of order $2 M_\odot$. 

Nevertheless,  it would be legitimate to argue that such a result could be an artifact of the model approximation adopted in Ref. \cite {Zhao:2019xqy}. After all, when compared to  QCD, the (effective) NJL model has some limitations which include  the absence of asymptotic freedom and non-renormalizability. On the other hand, the determination of the QCD EoS for cold and compressed matter, from {\it ab initio} evaluations, presents further issues which have not yet been circumvented. For instance, within this regime  lattice QCD (LQCD) simulations are still plagued by the infamous  sign problem \cite{deForcrand:2009zkb,Aarts:2015tyj} while perturbative QCD (pQCD) applications 
are only reliable at extremely high baryon densities, of order $n_B \sim 40 n_0$
\cite{Kurkela:2009gj,Fraga:2013qra,Fraga:2015xha,Annala:2017llu} ($n_0 = 0.16 \,
{\rm fm}^{-3}$), where asymptotic freedom allows for  weak coupling expansions.
Regarding pQCD it is important to mention that when the modified minimal
subtraction ($\overline {\rm MS}$) renormalization scale ($\Lambda$) is taken at
the conventional “central” value, $2 (\mu_u +\mu_d+\mu_s)/3$, with $\mu_f$ the
quark flavor chemical potentials, the next-to-leading order NLO pQCD  predicts
QS masses above $2 M_\odot$ \cite{Fraga:2004gz} and below $2 M_\odot$ at NNLO
\cite{Fraga2001,Fraga:2001xc,Kurkela:2009gj,Jimenez:2019iuc,Jimenez:2021wik}.

Although the cold dense pQCD NNLO pressure is formally perturbatively renormalization group (RG) invariant, the higher order residual scale dependence 
of pQCD results \cite{Kurkela:2009gj,Fraga:2013qra,Fraga:2015xha} is still rather sizeable, conventionally taken to lie between half and twice the central scale choice value. 
Quite recently, pQCD calculations reached 
NNNLO \cite{Gorda:2018gpy,Gorda:2021znl,Gorda:2021kme,Gorda:2023mkk} within the massless quark approximation, and all orders in the leading soft logarithms approximation \cite{Fernandez:2021jfr}, producing in such cases significant residual scale dependence reductions. 
However, accounting for the strange quark mass effects, sizeable at moderate and low $\mu$ values, is very involved and presently limited to NNLO, thus impiging on scale uncertainties reductions. In order to tackle the scale dependence problem an alternative method, which better incorporates RG properties within (variationally) optimized perturbative evaluations has been proposed \cite{Kneur:2013coa}. 
This {\it renormalization group optimized perturbation theory} (RGOPT) technique has been successfully  used, at vanishing temperatures and densities, to calculate quantities such as the QCD scale $\Lambda_{\overline{\rm MS}}$ 
and strong coupling $\alpha_s$ \cite{Kneur:2013coa}, and the quark condensate up to four and five loop orders \cite{Kneur:2015dda,Kneur:2020bph}.   
Later, the RGOPT has been applied to scalar field theories at finite temperatures \cite{Kneur:2015moa,Kneur:2015uha,Ferrari:2017pzt,Fernandez:2021sgr}, indicating that  the method greatly improves the scale dependence in thermal theories, when compared to {\it screened perturbation theory} (SPT) \cite{Karsch1997,Andersen:2000yj} and {\it hard thermal loop perturbation theory} (HTLpt) \cite{Andersen:1999fw,Andersen:2000yj,Andersen:2002ey} based on
the HTL effective theory \cite{Braaten:1991gm}. 
The technique was also employed to describe cold and dense QCD \cite{Kneur:2019tao}, as
well as the quark sector of QCD at finite temperatures and baryonic densities \cite{Kneur:2021dfo,Kneur:2021feo}. 
Once again, the RGOPT predictions display a residual scale dependence which is smaller  than those furnished by pQCD and HTLpt.
Recently, some of the present authors have used the RGOPT  EoS for QCD in the
chiral limit ($m_u=m_d\equiv 0$) to describe non-strange quark stars (NSQS)
obtaining encouraging results at NLO \cite {Restrepo:2022wqn}. In particular,
when the renormalization scale is appropriately chosen, the RGOPT resummation of
the perturbative series is able to describe several observational data,
specially  from pulsars PSR J0030+0451 \cite{Riley_2019,Miller:2019cac}, PSR
J0740+6620 \cite{Riley:2021pdl,Miller:2021qha,Cromartie:2019kug} and the light
compact object  identified in the middle of a supernova remnant, HESS J1731-347
\cite{2022NatAs.tmp..224D}.
Note that, recently a different approach has
been considered to constrain the scale $X$ within a model, possessing a bag
parameter constrained by  LQCD, which describes an isospin-dense matter system
\cite{Bai:2024amm}.
 
Here, our major goal is to investigate how the inclusion of strangeness  affects the $N_f=2$ results obtained in Ref. \cite {Restrepo:2022wqn}. In the present $N_f=2+1$ case, the introduction of a finite (strange) quark mass ($m_s$) modifies the RG operator introducing  an extra scale dependence, through the running of $m_s(\Lambda)$, which needs to be taken into account by the resummation procedure. 
This is accordingly a nontrivial extension with respect to the determination of the RGOPT EoS for massless quarks at $N_f=3$ \cite {Kneur:2019tao,Kneur:2021dfo,Kneur:2021feo} and $N_f=2$ \cite {Restrepo:2022wqn}.
Following the usual procedure, here we will also choose the scale to be density dependent so that it can be written in terms of the baryonic chemical potential\footnote {In this situation the pressure needs to be modified to ensure thermodynamic consistency.} ($\mu_B$) as $\Lambda = X \mu_B/3$, where $X$ represents a free parameter that can be adjusted in order to reproduce the desired  maximum mass value for a given  family of stars. Then, the QS properties determined from   the  novel   $N_f=2+1$ RGOPT EoS will be compared with those previously obtained with the same method when $N_f=2$ \cite{Restrepo:2022wqn} as well as with those furnished by $N_f=2+1$ pQCD at NNLO \cite{Kurkela:2009gj,Fraga:2013qra}, for the corresponding flavor species. 
We stress that our approach is not a fixed order perturbative
calculation but actually resumming a certain class of higher order contributions, indeed
in the massless quark approximation the NLO RGOPT pressure is 
definitely numerically closer to NNLO pQCD than to NLO pQCD~\cite{Kneur:2019tao}, and the latter
would be a poor approximation to consider compared to the state-of-the art NNLO pQCD. 
Therefore, in the present case with $m_s \ne 0$ we also consider appropriate to compare NLO RGOPT 
with NNLO pQCD.
Since $\alpha_s$ runs with $\Lambda =  X \mu_B/3$ the value of $X$ needed to produce a massive quark star will enable us to compare the coupling strength associated with each different number of flavors. This comparison will then allow us to  draw further conclusions about the reliability of the distinct EoS provided by the RGOPT at NLO.

The work is organized as follows. In the next section we present the perturbative result for QCD at NLO considering the case of quarks with a generic finite mass parameter. The RGOPT resummation setup is presented in Sec III. In Sec IV we discuss the modifications required to assure thermodynamic consistency as well as charge neutrality and $\beta$-equilibrium. The RGOPT and pQCD results for $N_f=2$ and $N_f=2+1$ are compared in Sec. V while Sec. VI contains our final conclusions.

\section{Perturbative QCD pressure for $N_f=2+1$ at NLO }
Let us start by reviewing the complete order-$\alpha_s$ massive quark perturbative pressure, at vanishing temperatures and finite chemical potentials, in the case of symmetric matter ($\mu_d=\mu_u=\mu_s\equiv \mu$). By adding the massive vacuum ($\mu=0$) results from Ref. \cite{Kneur:2015dda} to the in-medium ($\mu\ne 0$) results from Refs. \cite{Akhiezer:444287,Farhi:1984qu,Fraga:2004gz} one can write the   {\it per flavor} pressure, at NLO, as \cite{Kneur:2019tao}
 \begin{align}
\begin{split}
 P^{\rm{PT}}_{f}(m_f,\mu)=&-N_c \frac{m_f^4}{8 \pi^2} \left(\frac{3}{4}-L_m\right)+\Theta(\mu^2-m_f^2)\,\frac{N_c}{12 \pi^2} 
\left [ \mu p_F\left (\mu^2 - \frac{5}{2} m_f^2 \right ) + \frac{3}{2} m_f^4 \ln ( \frac{\mu +p_F}{m_f} ) 
\right ] \\
 &- \frac{g^2 d_A}{4\left(2\pi\right)^4} m_f^4 \left(3 L_m^2-4 L_m+\frac{9}{4}\right) 
 -\Theta(\mu^2-m_f^2)\, \frac{g^2 d_A}{4\left(2\pi\right)^4} \left \{ 3 \left [m_f^2 \ln ( \frac{\mu +p_F}{m_f} ) 
 -\mu p_F\right ]^2 - 2 p_F^4 \right \}
  \\
&- \Theta(\mu^2-m_f^2)\, \frac{g^2 d_A}{4\left(2\pi\right)^4} m_f^2  \left ( 4- 6 L_m \right ) 
 \left [ \mu p_F - m_f^2 \ln ( \frac{\mu +p_F}{m_f} ) \right ] \,,
\label{PPTqcd} 
\end{split}
\end{align}
where $p_F=(\mu^2-m_f^2)^{1/2}$ is the Fermi momentum, $L_m=\ln(m_f/\Lambda)$, $d_A=N_c^2-1$, $\Lambda$ is the $\overline{\rm MS}$ arbitrary 
renormalization scale, while $\Theta$ represents the 
Heaviside function and $g^2\equiv 4\pi\alpha_s$.
Being independent of the medium, the pure vacuum contributions $\propto m^4$, although originally present in the basic calculation, are often simply discarded in the literature. However, since renormalization properties essentially depend on vacuum contributions, the latter actually play a crucial role in our construction based on RG properties, as will be clear below.
It is often useful to further simplify Eq. (\ref{PPTqcd}) in order to take the relevant case $\mu>m$ into account.  It then reads  
\begin{align}
\begin{split}
 P^{\text{PT}}_{f}(m_f,\mu)=& 
  \frac{N_c}{12 \pi^2} \left [ \mu p_F\left (\mu^2 - \frac{5}{2} m_f^2 \right ) 
  + \frac{3}{2} m_f^4 \left ( L_\mu -\frac{3}{4} \right ) \right ]\\
  &-\frac{g^2 d_A}{4\left(2\pi\right)^4}\left[ m_f^4 \left ( 3 L_\mu^2- 4 L_\mu + \frac{1}{4} \right ) +
 \mu^2 \left(\mu^2+m_f^2 \right) +m_f^2\,\mu \, p_F \left(4 - 6 L_\mu \right) \right]  ,
\end{split}
\label{PPTqcdsimp}
 \end{align}
where 
\begin{equation}
L_\mu \equiv \ln [(\mu+p_F)/\Lambda]\; .
\label{Lmudef}
\end{equation}
Note, however, that Eqs. (\ref{PPTqcd}), (\ref{PPTqcdsimp}) are not perturbatively RG invariant, since the renormalization scale $ \ln(\Lambda)$ enters the LO ${\cal O}(g^0)$ contribution. This is inherently
related to the extra vacuum energy UV divergences of any massive theory, but perturbative RG invariance can be consistently recovered as specified next.
\section{RGOPT resummation setup}\label{secIII}

Before applying the RGOPT prescription one must  obtain a (perturbatively) RG invariant (RGI) pressure,  which can be achieved  by adding additional zero-point contributions to the pressure given by Eq. (\ref{PPTqcd}). The latter prescription can then be written as \cite{Kneur:2015dda,Kneur:2019tao}
\begin{align}
 P^{\rm{PT}}_{f}(m_f,\mu) \to P^{\rm{RGI}}_{f}(m_f,\mu)=P^{\rm{PT}}_{f}(m_f,\mu) -m_f^4 \sum_k s_k g^{2k-2}\; ,
 \label{sub}
\end{align}
so that the resulting pressure becomes scale independent at a given perturbative order,
up to neglected higher orders. The coefficients $s_k$ are determined by applying the 
massive RG operator\footnote{Note that the $s_i$ coefficients are also related to the 
so-called vacuum energy anomalous dimension coefficients \cite{Chetyrkin:1994ex,Baikov:2018nzi}, as discussed in more
details e.g. in Ref. \cite{Kneur:2020bph}.}
\begin{align}
 \Lambda \frac{d}{d \Lambda}= \Lambda \frac{\partial}{\partial \Lambda} + \beta(g^2 )\frac{\partial}{\partial g^2 } - 
\gamma_m(g^2 ) m_f \frac{\partial}{\partial m_f} \;,
\label{RG}
\end{align}
at successive perturbative orders. One then obtains \cite{Kneur:2019tao} 
\begin{equation}\label{s0def}
 s_0=- N_c \left[(4\pi)^2 (b_0-2\gamma_0)\right]^{-1} \;,
 \end{equation}
 and
 \begin{equation}
 s_1=-\frac{N_c}{4}\left[\frac{b_1-2\gamma_1}{4(b_0-2\gamma_0)} -\frac{1}{12\pi^2}\right ]\,.
 \label{s1}
\end{equation}
In our notation, the coefficients of the $\beta$ and $\gamma_m$ RG functions are given by 
\begin{equation}
 \beta\left(g^2\equiv 4\pi\alpha_s \right)=-2b_0g^4-2b_1g^6+\mathcal{O}\left(g^8\right) \;,
 \label{betaQCD}
 \end{equation}
 and
 \begin{equation}
 \gamma_m\left(g^2\right)=\gamma_0g^2+\gamma_1g^4+\mathcal{O}\left(g^6\right)\;,
\end{equation}
where 
\begin{align}
 b_0=& \frac{1}{\left(4\pi\right)^2}\left(11-\frac{2}{3}N_f\right ),\\
 b_1=& \frac{1}{\left(4\pi\right)^4}\left (102-\frac{38}{3}N_f\right ),\\
 \gamma_0=&\frac{1}{2\pi^2},
 \end{align}
 and
\begin{equation}
 \gamma_1^{\overline {\rm MS}}= \frac{1}{8\left(2\pi\right)^4}\left (\frac{202}{3}-\frac{20}{9}N_f\right) \;.
\end{equation}
Generally, the RGOPT pressure ($P^{\rm RGOPT}$) is then obtained by implementing the following steps (see, e.g., Refs. \cite{Kneur:2013coa,Kneur:2019tao} for a review):\\
\begin{enumerate}
\item First, one modifies the free and interaction terms in the (renormalized) Lagrangian density by performing  the replacements
\begin{equation}
 m_f\to m_f+\eta\left(1-\delta\right)^a, \;\;  g^2\to \delta g^2,
 \label{delta}
\end{equation}
most conveniently within the renormalized  RGI massive pressure, Eq. (\ref{sub}). 
In Eq. (\ref{delta}) $\delta$ is the new expansion parameter and the exponent $a$ is specified below.
In QCD applications, $m_f$ represents the physical (current) quark mass, while $\eta$ represents an {\it arbitrary} mass parameter, whose optimal value is subsequently fixed by RG properties and a variational stationary criterion, as will be specified below. This will generate a RG-dressed screening quark mass, $\eta(g^2,\mu)$, with a resummed coupling dependence due to RG resummation properties.
Note that, for $a=1$, Eq. (\ref{delta}) is equivalent to the more familiar ``added and subtracted" mass term prescription typically adopted in SPT \cite{Andersen:2000yj} or HTLpt \cite{Andersen:2002ey,Haque:2012my}.
\item Next, one expands the pressure resulting from Eq. (\ref{delta}) to the desired 
(modified) perturbative order in powers of  $\delta$,  which 
is set to the unit value in order to recover the original theory at the end of calculations. 
However, part of the RGOPT procedure stems from the observation that Eq. (\ref{delta}) may generally spoil the perturbative  RGI properties of the original pressure, Eq. (\ref{sub}) (even in the chiral limit, $m_f\to 0$). 
Therefore, the exponent $a$ in Eq. (\ref{delta}) is instead (uniquely) fixed by the RG equation at LO in the massless case, $m_f=0$, 
to the critical value
\begin{equation}\label{acrit}
a= \frac{\gamma_0}{2b_0}\;,
\end{equation}
so that the RG invariance is recovered (for $m_f\to 0$) at LO, as prescribed in previous RGOPT applications \cite{Kneur:2013coa,Kneur:2019tao}.
Remark that, when $m_f=0$, the RG operator given by Eq. (\ref{RG})  reduces to 
\begin{align}
 \Lambda \frac{d}{d \Lambda}= \Lambda \frac{\partial}{\partial \Lambda} + \beta(g^2 )\frac{\partial}{\partial g^2 }  \;.
\label{RGred}
\end{align}
\item Obviously, the previous step  leaves a remnant $\eta$-dependence at any finite $\delta$ orders, 
similarly to the original OPT (or HTLpt \cite{Haque:2012my}). This initially arbitrary parameter  may be fixed   by requiring the pressure, obtained 
from the modified perturbative expansion, to satisfy a mass optimization prescription (MOP)  
\begin{equation}
f_{\rm MOP}\equiv     \frac{\partial P^{\rm RGOPT}}{\partial \eta} \Big \vert_{\overline \eta} = 0 \;,
 \label{pms}
\end{equation}
so that the optimal variational mass ($\overline \eta$) becomes a function of the {\it original} parameters (such as the couplings) as well as control parameters (like the chemical potential).
\item Although at LO and $m_f= 0$ the RG Eq. (\ref{RGred}) is automatically fulfilled due to Eq. (\ref{acrit}), at NLO and higher orders (or for $m_f\ne 0$) 
the resulting pressure no longer satisfies Eq. (\ref{RGred}),
due to reshuffled mass dependence. Thus, a possible alternative prescription to Eq. (\ref{pms}) is to (re)impose the RG Eq. (\ref{RGred}) or, more precisely, Eq. (\ref{RG})
for $m_f\equiv m_s \ne 0$
\footnote{Note in Eq. (\ref{fRGs}) the additional anomalous mass dimension consistently due to the physical mass
 $m_s$, on top of the massless RG operator.}
\begin{equation}\label{fRGs}
 f_{{\rm RG}} \equiv   
 \Lambda \frac{\partial P^{\rm{RGOPT}}_{{\rm NLO}}}{\partial \Lambda} + \beta(g^2 )\frac{\partial P^{\rm{RGOPT}}_{{\rm NLO}}}{\partial g^2 } - 
\gamma_m(g^2 ) m_s \frac{\partial P^{\rm{RGOPT}}_{{\rm NLO}} }{\partial m_s} =0\,. 
 \end{equation}
\end{enumerate}

At this point, some important additional remarks are in order: 
First, note that although these RG properties were originally obtained from vacuum ($\mu=T=0$) contributions, they still hold when thermal and/or in-medium 
effects are considered \cite{Kneur:2015uha,Kneur:2015moa,Ferrari:2017pzt,Kneur:2021dfo,Kneur:2021feo}.
Also, according to the original prescription, $a$ is fixed once for all at LO, and the same value 
in Eq. (\ref{acrit}) is used
at successive perturbative orders of the (modified) perturbative $\delta$-expansion. 
Importantly, this guarantees \cite{Kneur:2013coa} in particular that at higher orders there is always
one solution $\overline \eta(g)$ compatible with asymptotic freedom behavior for $g\to 0$.
Finally, note that even when treating massive theories as we do here, it is compelling to adopt this universal constant value for $a$ in Eq. (\ref{acrit}), in order to avoid otherwise 
a clumsy $a(g, m,...)$ dependent exponent, furthermore incompatible with standard renormalization. We emphasize that although other slightly different variational prescriptions could be thought of, 
the latter one provides a sensible way of comparing successive perturbative orders with the same
prescription. 
This is further justified a posteriori since 
the strange quark mass ($m_s$) which is  relevant here  remains a rather moderate perturbation\footnote{As we will see in the sequel, $m_s \ll \overline \eta \simeq {\cal O}(g\, \mu) \ll \mu$ roughly holds, at least as long as the QCD coupling, $g^2(\Lambda)$, remains moderate (i.e.
for renormalization scales $\Lambda \sim {\cal O}(\mu)$ with $\mu$ not too small).}
with respect to the dressed mass $\eta(g^2,\mu)$ determined from Eq. (\ref{pms}). 

One can next apply the RGOPT replacements to the RGI pressure, Eq. (\ref{sub}), 
to similarly obtain the NLO pressure. Therefore, the pressure corresponding to the  strange sector can be written 
as\footnote{Note that Eqs. (\ref{rgoptNLO}), (\ref{rgoptLO}) are not simply obtained
from replacing $m_f\to m_s+\eta$ within the original NLO expression Eq. (\ref{PPTqcdsimp}), due to the modified perturbative expansion from Eq. (\ref{delta}).}
%
\begin{align}
\begin{split}
 P^{\rm{RGOPT}}_{{\rm NLO},s}(m_s,\eta,\mu)=&   P^{\rm{RGOPT}}_{{\rm LO},s}(m_s,\eta,\mu) - N_c \frac{\eta\left(\eta+m_s\right)^2}{\left(4\pi\right)^2 b_0\,g^2} \left(\frac{\gamma_0}{b_0}\right)  \left ( \eta-\frac{\gamma_0-2b_0}{2\left(b_0-2\gamma_0\right)}m_s \right)  \\
&-N_c \frac{\left(\eta+m_s\right)^3}{4} \left [\eta\left(2\frac{\gamma_0}{b_0} -1\right)+m_s\right ] \left(\frac{b_1-2\gamma_1}{4(b_0-2\gamma_0)} -\frac{1}{12\pi^2}\right )\\
&+ N_c \frac{\eta\left(\eta+m_s\right)}{8\pi^2} \left( \frac{\gamma_0}{b_0}\right ) 
\left [ \left(\eta+m_s\right)^2 \left( 1 - 2L_{\mu,s} \right ) +2 \mu \, p_{F,s}  \right ]  \\ 
 &- \frac{g^2 d_A} {4\left(2\pi\right)^4} \left[ \left(\eta+m_s\right)^4 \left (\frac{1}{4} - 4 L_{\mu,s} +3 L_{\mu,s}^2 \right ) +
 \mu^2 \left(\mu^2+\left(\eta+m_s\right)^2 \right) \right .\\
 &+ \left . \mu  p_{F,s}\left(\eta+m_s\right)^2 \left(4 - 6 L_{\mu,s} \right) \right]   \;,
\end{split} \label{rgoptNLO}
\end{align}
where the LO result reads
\begin{align}
\begin{split}
  P^{\rm{RGOPT}}_{{\rm LO},s}(m_s,\eta,\mu)=&
  \frac{N_c}{12 \pi^2} \left [ \mu p_{F,s}\left (\mu^2 - \frac{5}{2} \left(\eta+m_s\right)^2 \right ) 
  + \frac{3}{2} \left(\eta+m_s\right)^4 \left ( L_{\mu,s} -\frac{3}{4} \right ) \right ] \\
  & + N_c \frac{\left(\eta+m_s\right)^3}{\left(4\pi\right)^2 b_0 \, g^2}\left(\eta +\frac{b_0}{b_0-2\gamma_0} m_s\right)\;,
\end{split} \label{rgoptLO} 
\end{align}
and $p_{F,s}=[\mu^2-(\eta+m_s)^2]^{1/2}$, so that $L_{\mu,s} = \ln [(\mu+p_{F,s})/\Lambda]$ changes accordingly. 

To consider the sector corresponding to the pair of light  quarks one just needs to take $m_s\to m_u=m_d\equiv 0$  in Eq. (\ref{rgoptNLO}). Then, upon considering Eq. (\ref{rgoptNLO})  one can  write  the total RGOPT pressure at $N_f=2+1$ as 
\begin{equation}
P^{\rm{RGOPT}}_{{\rm NLO}}(m_s,\eta,\mu)= P^{\rm{RGOPT}}_{{\rm NLO},s}(m_s,\eta,\mu) + 2 P^{\rm{RGOPT}}_{{\rm NLO},s}(0,\eta,\mu)\,.
\label{Ptot21}
\end{equation}
Finally, to make the expressions self-contained, let us also remark that when considering the approximation of $N_f$  massless physical quarks, the result is just 
\begin{equation}
P^{\rm{RGOPT}}_{{\rm NLO}}(\eta,\mu)=  N_f P^{\rm{RGOPT}}_{{\rm NLO},s}(0,\eta,\mu)\,,
\label{Ptot3}
\end{equation}
in accordance with Refs. \cite{Kneur:2019tao,Restrepo:2022wqn}. By comparing the results provided by Eqs. (\ref{Ptot21}) and (\ref{Ptot3}) we will be able to gauge  how the presence of a realistic running quark mass, on top of the RG-dressed screening mass $\overline\eta(g^2,\mu)$, affects the RGOPT predictions.
\subsection{Optimization}
As discussed above and in previous applications to QCD \cite{Kneur:2019tao,Kneur:2021dfo,Kneur:2021feo}, concerning the NLO RGOPT pressure one may either consider the dressed mass $\overline \eta(g^2,\mu)$  obtained from solving 
the MOP Eq. (\ref{pms}) or, 
alternatively, the dressed mass  obtained from the {\it reduced} RG relation, Eq. (\ref{fRGs}). However, our  previous RGOPT application \cite {Kneur:2019tao} to the $N_f=3$ case (which is equivalent to the  $m_s\to 0$ approximation within the $N_f=2+1$ results considered here) indicates that at NLO, neither 
the MOP Eq. (\ref{pms}) nor the RG Eq. (\ref{RGred}), when solved exactly,
produce real-valued $\overline\eta(g^2,\mu)$ solutions over the full range of phenomenologically relevant values of the chemical potentials. This is
due to a very nonlinear dependence in  the (variational) mass\footnote{On the other hand, more perturbative (reexpanded) approximate real solutions 
could be considered, but at the price of loosing most of the built-in higher order RG invariance properties of $\overline\eta(g^2,\mu)$.}. 
This issue can be remediated, however, upon performing a perturbative renormalization scheme change (RSC) 
such that a real solution is recovered \cite{Kneur:2013coa}. 
In practice the RSC can be simply implemented at the relevant NLO by modifying the (variational) mass parameter which appears in the RGI pressure, Eq. (\ref{sub}) (that is, prior to the RGOPT prescription in Eq. (\ref{delta}), such that the RSC is perturbatively well defined). Such modification can be carried out according to \cite{Kneur:2013coa,Kneur:2019tao}
\begin{align}
 m_f\to m_f (1+ g^4 B_2 )\,.
 \label{RSC}
\end{align}
After applying this RSC to the original NLO pressure, one must re-expand 
perturbatively the latter to NLO, producing an extra term 
$-4g^2\,m_f^4s_0 B_{2}$ in the resulting pressure. As a by-product, after the RGOPT modifications, similar contributions
with $m_f\to \eta +m_s $ and  $m_f\to \eta $ are induced respectively in the first and second term on the right hand side of Eq. (\ref{Ptot21}).
 Although the dimensionless RSC parameter $B_2$ in Eq.(\ref{RSC}) is
arbitrary to start with, one seeks a
 sensible prescription to recover real solutions while maintaining $B_{2}$ 
reasonably close to $\overline{\rm MS}$ results to justify a posteriori such a  perturbative RSC. As Ref. \cite{Kneur:2013coa} prescribes, 
$B_2$ is accordingly fixed (uniquely) upon requiring the nearest ``contact'' of the two curves 
parametrizing the MOP Eq. (\ref{pms}) and RG Eq. (\ref{RGred}), considered
as functions $f_{\rm{MOP}}(\eta, g^2)$, $f_{\rm{RG}}(\eta, g^2)$ respectively. 
Algebraically this corresponds to solving 
\begin{equation}
f_{\rm RSC} = \frac{ \partial f_{\rm RG}}{\partial g^2}\frac{ \partial f_{\rm MOP}}{\partial \eta} - \frac{ \partial f_{\rm RG}}{\partial \eta}\frac{ \partial f_{\rm MOP}}{\partial g^2} \equiv 0\,,
\label{RSC_eq}
\end{equation}
simultaneously with Eq. (\ref{pms}), $f_{\rm{MOP}}(g^2,m)=0$, in order to obtain the two (real)  values $\overline \eta$, ${\overline B}_2$ 
(for the $N_f=3$ case as well as for the  $N_f=2+1$ case at hand). 

For the running coupling $g^2(\Lambda)$, we use the exact two-loop order, obtained from 
\begin{align}
 \ln \frac{\Lambda}{ \Lambda_{\overline{\text{MS}}} } = \frac{1}{2b_0\, g^2} +
\frac{b_1}{2b_0^2} \ln \left ( \frac{b_0 g^2} {1+\frac{b_1}{b_0} g^2} \right) ,
\label{g2L}
\end{align}
where $\Lambda_{\overline{\rm{MS}}}\simeq 0.335\,{\rm GeV} $ ($N_f=3$) is fixed so that $\alpha_s(\Lambda = 1.5 \, {\rm GeV})\simeq 0.326$ \cite{Bazavov:2012ka}, in rough consistency with the latest world-average values \cite{ParticleDataGroup:2018ovx}.
For the strange quark  mass we consider the NLO running, 
\begin{align}
    m_s=\hat{m}_s\left(\frac{g^2}{4\pi^2}\right)^{4/9}\left(1+0.895062\frac{g^2}{4\pi^2}\right), \label{msrunning}
\end{align}
with $\hat{m}_s\simeq 0.253$ GeV corresponding to $m_s(2\, \rm{GeV})=0.0935(8)\, \rm {GeV}$ \cite{ParticleDataGroup:2018ovx}. 
\begin{figure}
\begin{subfigure}
  \centering
 \includegraphics[width=.45\textwidth]{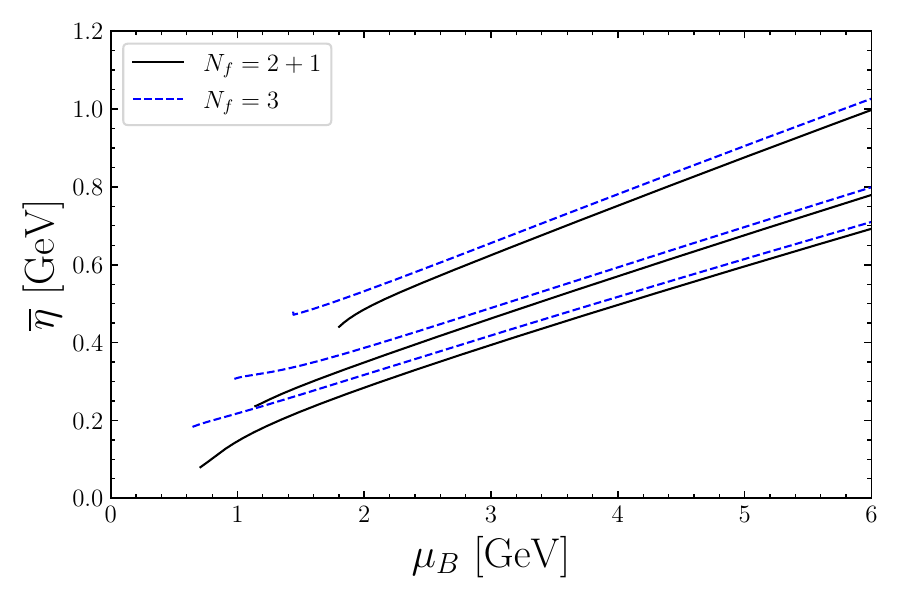}
\end{subfigure} 
\begin{subfigure}
  \centering
 \includegraphics[width=.45\textwidth]{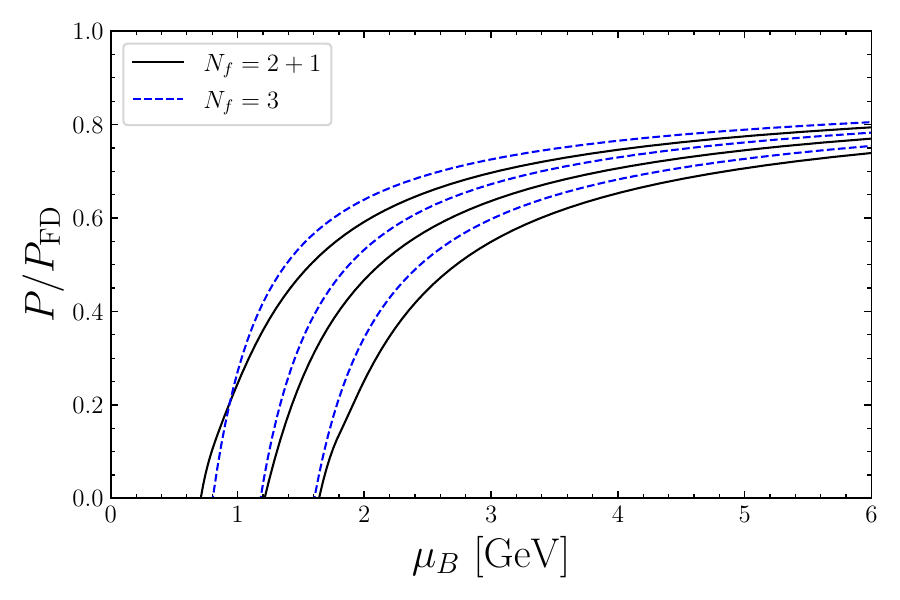}
\end{subfigure} 
\caption[large]{Left panel: NLO optimized mass parameters, $\overline{\eta}$ as functions of the baryon chemical potential for $N_f=2+1$ ($m_s\ne 0$) (continuous line) and $N_f=3$ ($m_s=0$) (dashed lines). 
Upper lines correspond to $X=1$, lower lines to $X=4$ and central lines correspond to $X=2$.
 Right panel: Normalized pressure as functions of the baryon chemical potential for $N_f=2+1$ and $N_f=3$. The bottom lines correspond to $X=1$, 
 the upper lines to  $X=4$ while the central lines correspond  to $X=2$.} 
\label{eta_B2}
\end{figure}

Fig. \ref {eta_B2} compares the optimized mass parameter,  $\overline{\eta}$, as well as the normalized pressure as functions of the baryon chemical potential representing symmetric matter, $\mu_B=3\mu$, 
for $N_f=3$   and $N_f=2+1$. 
 Following the  standard prescription, we have considered three different values of the renormalization scale, $\Lambda=X\mu_B/3$, with $X=1,2,4$, to quantify the residual scale dependence.
Regarding the variational masses, the figure suggests that they share a similar behavior, increasing  linearly with $\mu_B$ roughly as $\overline \eta\sim {\cal O}(g\,\mu)$, although
 $\overline \eta$ actually entails an all-order $g^2$-dependence from RG resummation. Importantly, the fact that $\overline \eta$ accordingly behaves as a screening mass
 for moderate coupling $g$, also guarantees that the RSC Eq. (\ref{RSC}) modifications induced to the pressure,
 $\propto -B_2 s_0 g^2 \overline \eta^4$, are of formally higher orders ${\cal
O}(g^6)$, so that
 the physical pressure is only mildly dependent on the renormalization scheme
choice. For the same  $\mu_B$ value, and comparing with both $N_f=2$ and $N_f=3$
cases, the presence of a massive quark ($N_f=2+1$) generates a lower pressure.
This can be explained by first noting that
 in the basic NLO pressure Eq. (\ref{PPTqcd}) or (\ref{PPTqcdsimp}), the $m_f \ne 0$ 
 contributions are always negative
 as compared to the equivalent massless contributions (at least as long as $m_f \ll \mu$), 
 and this feature remains true for the modified NLO expression Eq. (\ref{rgoptNLO}). 
 Then since $\overline\eta +m_s > \overline\eta$, the lowering of the pressure is obviously
 more pronounced for $m_s \ne 0$ in the RGOPT 
 case\footnote{In contrast, the RSC modification $\sim -B_2 s_0 g^2 \overline \eta^4$ gives
 a {\it positive} extra contribution to the pressure, as $s_0 >0$ in Eq. (\ref{s0def}) and we always obtain $B_2 <0$. But this contribution
 is subdominant compared to the "direct" $\overline\eta$ dependence entering other
 contributions.}. A lower pressure tends to enhance the remnant scale dependence, 
 therefore slightly counteracting the scale dependence improvement obtained from RGOPT, although the figure indicates that the observed scale dependence is not so much sensitive to the strange quark mass value.

\section{Chemical equilibrium and thermodynamic consistency}
Before performing numerical evaluations to further analyze the relevant physical observables, let us redefine some thermodynamic quantities to account for chemical equilibrium and to ensure thermodynamic consistency. The latter requirement is important if one recalls that the renormalization scale (and hence the coupling) is  taken to be density dependent.
When considering the $N_f=2+1$ case one can enforce chemical equilibrium and charge neutrality   by requiring the following  relations to be satisfied
\begin{align}
 \mu_u=\mu_d-\mu_e\equiv \mu, \ \ \ \ \mu_s=u_d, \label{betaeq_nf2}\\
 \frac{2}{3}\rho_u-\frac{1}{3}\rho_d-\frac{1}{3}\rho_s-\rho_e=0\,.
 \label{charge_neu_nf3} 
\end{align}
When $N_f=2$, the similar relations 
\begin{align}
 \mu_u=\mu_d-\mu_e\equiv \mu, \label{betaeq_nf3}\\
 \frac{2}{3}\rho_u-\frac{1}{3}\rho_d-\rho_e=0\,,\label{charge_neu_nf2} 
\end{align}
must be observed.
The relevant baryon chemical potential describing each different number of flavors is defined as $\mu_B=\mu_u+2\mu_d$ (recall that $\mu_d = \mu_s$). 
Seeking   a fair comparison, the renormalization scale given by $\Lambda=X\,\mu_B/3$ 
is chosen in all cases.\\ 
At this stage we should specify how to best compare our RGOPT pressure
results with pQCD,
which for the full $m_s$ dependence is known at NNLO~\cite{Kurkela:2009gj,Fraga:2013qra,Fernandez:2024ilg}. 
As anticipated in the introduction we consider more appropriate to compare NLO RGOPT with NNLO pQCD, 
since the NLO RGOPT expression given by  Eq.(\ref{rgoptNLO}) resums an all order RG dependence, and in the massless quark approximation the NLO RGOPT pressure is numerically closer~\cite{Kneur:2019tao} 
to NNLO pQCD than to NLO pQCD (we have also explicitly  checked
that a similar behavior holds for $m_s \ne 0$).
To obtain the pQCD results at $N_f=2$, we thus use the NNLO result for massless quarks given by \cite{Fraga2001,Fraga:2001xc,Vuorinen:2003fs},
 \begin{align}
\begin{split}
\frac{P^{\rm{pQCD}}_{N_f=2}}{P_{\rm{FD}}} =& 1 - 2\frac{g^2}{4\pi^2}  -\frac{g^4}{\left(4\pi\right)^2} \left [ 10.3754- 0.535832 N_f+N_f\ln(N_f)+ N_f \ln (\frac{g}{4\pi^2})+ \left(11-\frac{2}{3}N_f\right) \ln \left (\frac{\Lambda}{\mu_f} \right ) \right]   \; ,
 \label{pQCD_Nf2}
 \end{split}
\end{align}
where the Fermi-Dirac pressure $P_{FD}$ describing non-interacting massless quarks reads
\begin{equation}
 P_{\rm{FD}} = \frac{ 3 N_f } {12\pi^2}\left(\frac{\mu_B}{3}\right)^4 \;.  \end{equation}
Concerning the pQCD results at $N_f=2+1$, we consider the pocket formula given in Ref. \cite{Fraga:2013qra}, which reproduces well the NNLO results of Ref. \cite{Kurkela:2009gj}. This simple relation is  given by
\begin{align}
\begin{split}
     \frac{P^{\rm pQCD}_{N_f=2+1}}{P_{\rm{FD}}}=&\left(c_1-\frac{a(X)}{(\mu_B/{\rm} GeV)-b(X)}\right),\\
     a(X)=&d1X^{-\nu_1}, \ \ b(X)=d2X^{-\nu_2},
\end{split}
\label{pQCD_Nf3}
\end{align}
where $c_1=0.9008, d_1=0.5034, d_2=1.452,\nu_1=0.3553,$ and  $\nu_2=0.9101$.

As already emphasized, since the scale is chosen to be density dependent, the equations of state, Eqs (\ref{Ptot21}),(\ref{Ptot3}) and (\ref{pQCD_Nf2}), need to be redefined to fulfill thermodynamic consistency. The pQCD EoS given by Eq. (\ref{pQCD_Nf3}) is already thermodynamically consistent by construction \cite{Fraga:2013qra}. To achieve thermodynamic consistency within $N_f=2$ pQCD, 
and within the different $N_f$ cases considered for the RGOPT,  we follow the same procedure adopted in our previous work\footnote 
{At this point,  note a detail missed in Ref. \cite{Restrepo:2022wqn}: the electron density obtained from charge neutrality, Eqs. (\ref{charge_neu_nf3}) and (\ref{charge_neu_nf2}), also depends on the parameters $w_i$. Consequently, the electron pressure is not thermodynamic consistent. Then, in principle, one also needs to add a term $b_e$ to the leptonic pressure to completely guarantee thermodynamic consistency. For the sake of rigor, we have added this term in our present results. However, it turns out that the leptonic term, missed in Ref. \cite{Restrepo:2022wqn}, is small enough so that its contribution to the full pressure is almost negligible.} \cite{Restrepo:2022wqn}, where  extra terms $b_f$ are added to the pressure, defined so that 
\begin{align}\label{thcons}
    \frac{db_f}{d\mu_f}=-\sum_i \frac{dP_f}{dw_i}\frac{dw_i}{d\mu_f},
\end{align}
where $w_i$ represent Lagrangian parameters that depend on the control parameter $\mu_f$ (see Refs \cite{Gorenstein:1995vm,Restrepo:2022wqn, Lenzi:2010mz} for  a deeper discussion).
\section{Numerical Results}
In this section, we first present the numerical results furnished by the
RGOPT and pQCD for NSQM ($N_f=2$) as well as  SQM ($N_f=2+1$) considering
$\beta$-equilibrium, Eqs. (\ref{betaeq_nf2}) and (\ref{betaeq_nf3}), and charge
neutrality, Eqs. (\ref{charge_neu_nf2}) and (\ref{charge_neu_nf3}). Next,
the EoS corresponding to each case is  used to obtain the mass-radii relation
describing non-strange and strange quark stars.

\subsection{NSQM and SQM for $\beta$-equilibrium matter\label{beta}}

\begin{figure}
\begin{subfigure}
    \centering
 \includegraphics[width=.45\textwidth]{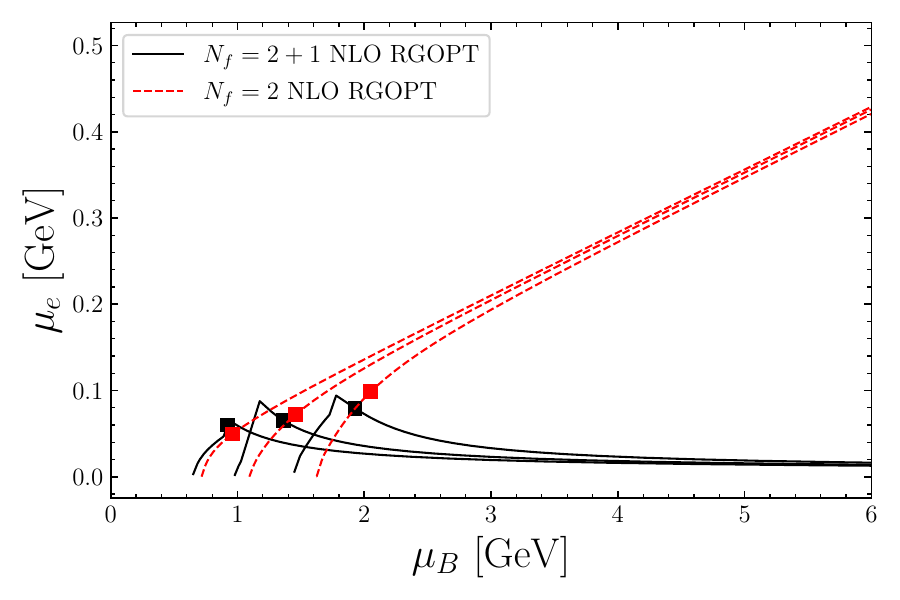} 
\end{subfigure}
 \begin{subfigure}
     \centering
     \includegraphics[width=.45\textwidth]{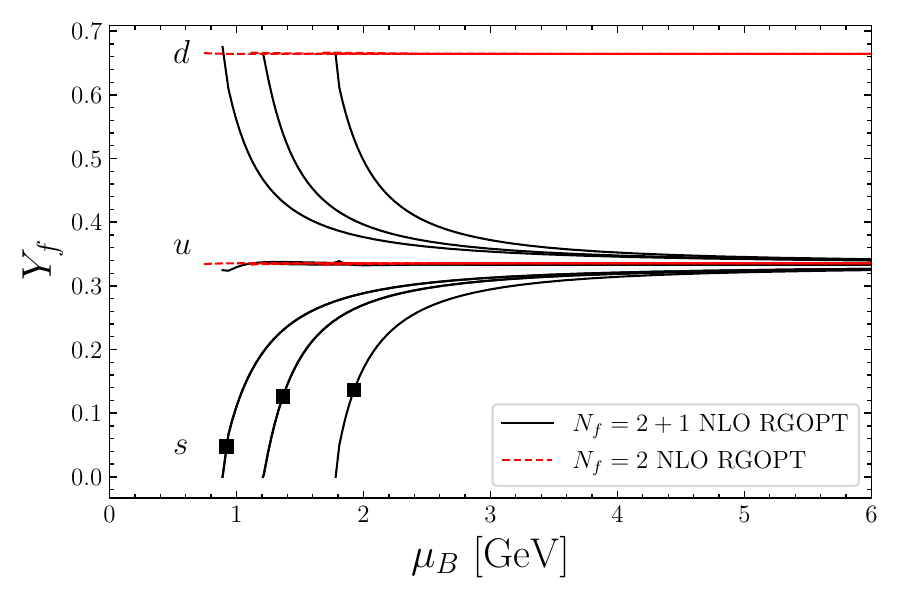}
 \end{subfigure}
\caption[long]{Left panel: electron chemical potential, as a function of $\mu_B$, obtained from the charge neutrality condition. The case $N_f=2+1$  is represented by the continuous line  while the dashed line represents the case $N_f=2$. The boundaries of each band are obtained by setting the renormalization scale coefficient to $X=1$ (rightmost boundary) and $X=4$ 
(leftmost boundary), while $X=2$ corresponds to the central line. The kink, in the case $N_f=2+1$, corresponds to the onset of the $s$ quark. Right panel: quark fraction $Y_f = \rho_f / \rho_B$ as a function of $\mu_B$. In both figures the squares represent the values at which $P=0$, which defines the surface of the QS.}
\label{muevsmu_B0}
\end{figure}

The electron chemical potential predicted by the NLO RGOPT is plotted  in Fig. \ref{muevsmu_B0} (on the left panel) as a function of $\mu_B$ for $N_f=2+1$ (solid lines) and $N_f=2$ (dashed lines). In the case of SQM, the appearance of kinks in the curves signals the onset of strange quarks which induces  a strong reduction of the number of electrons decreasing  the electron chemical potential ($\mu_e \to 0$, at high-$\mu_B$). Therefore, in this limit, the densities of the three quark flavors become similar. In other words, the system favors the replacement of the negative electric charge of electrons by the one of $s$-quarks. On the other hand, in the case of NSQM, as $\mu_B$ takes on higher values, the electron production increases to compensate for the increase in the $u$ quark number density, while maintaining charge neutrality.  {Note that the $u$ quark  has a smaller Fermi momentum than the $d$ quark and, therefore, an increase of the number density  is favored.} 

On the right panel of Fig. \ref{muevsmu_B0} the relative fraction of quarks  with different flavors, $Y_f = \rho_f/\rho_B$ is shown. At low chemical potential the $s$-quark has still not set in and the fraction of $d$-quarks is approximately equal to the double of the $u$-quark fraction due to charge neutrality. After the onset of strangeness,  $s$-quarks gradually replace  $d$-quarks while the yield of $u$-quarks does not change. At large chemical potentials, when the Fermi momenta are much larger than the effective masses, all fractions become similar. This is what is frequently designated by strange matter. Note, however, that the QS surface is defined by the condition $P=0$, and, therefore, the $s$-quarks are already present on that layer of the QS. 

\begin{figure}
  \centering
 \includegraphics[width=.6\textwidth]{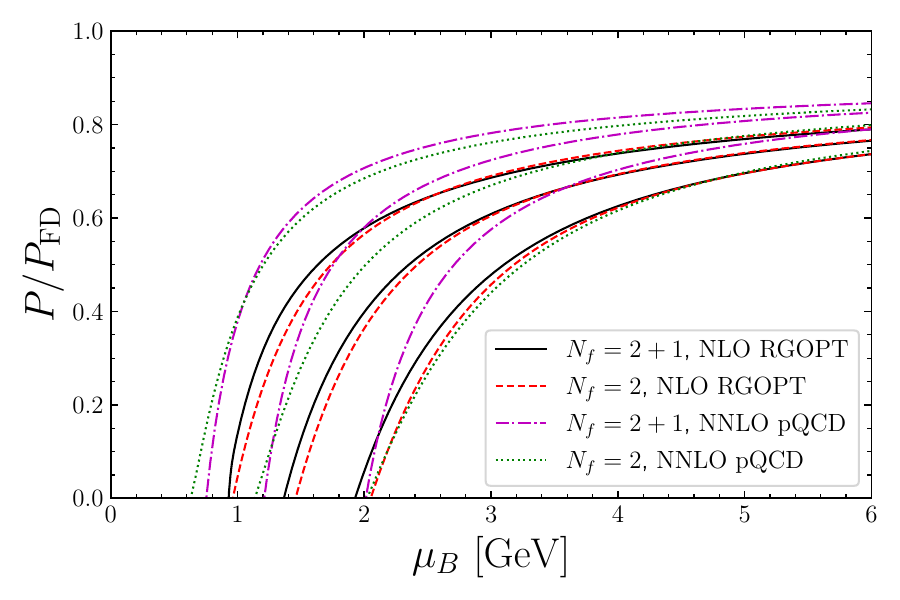}
\caption{ Thermodynamic consistent NLO RGOPT pressure as function of $\mu_B$ for $N_f=2+1$ (continuous lines) and  $N_f=2$ (dashed lines). For comparison the dot-dashed and dotted lines respectively represent the $N_f=2+1$ and $N_f=2$ NNLO pQCD pressure. 
For all curves the upper ones  correspond to $X=4$, the  lower ones to $X=1$ while the central scale, $X=2$, is represented by the inner curves. All pressures have been normalized by the Fermi-Dirac value $P_{\rm FD}$ for free massless quarks. }
\label{P_2L}
\end{figure}

In Fig. \ref{P_2L}, we show the normalized pressures furnished by the NLO RGOPT and NNLO pQCD for both SQM and NSQM, thermodynamically consistent after applying Eq. (\ref{thcons}). 
The first thing to note is that, within both approximations, the introduction of a finite strange quark mass decreases the scale dependence (this effect is much more noticeable in the case of pQCD).
While those features result from involved combined effects from the running coupling, running mass, and extra thermodynamic consistency
contributions, this behavior is explained partly from the fact that since $b_0(N_f=3)< b_0(N_f=2)$, the running coupling $g^2(\Lambda)\sim [2b_0 \ln (\Lambda/\Lambda_{QCD})\, ]^{-1}$ 
has somewhat smaller scale variation for $N_f=3$ than for $N_f=2$. The latter 
effect is however partly compensated by the sizable massive contributions $\overline\eta \ne 0$, lowering the pressure (as explained in the last paragraph before Sec. \ref{secIII}), which
rather tends to increase the remnant scale dependence.
As explained in Sec. III, the RGOPT procedure involves a resummed
all order RG dependence, thus adding higher order
$\Lambda$-dependence that can further reduce the
renormalization scale dependence, so that our results display a somewhat reduced scale dependence with respect to unresummed pQCD. However, the improvement that 
we obtain for the present cold quark matter case is much more moderate than its counterpart for hot
QCD in \cite{Kneur:2021dfo,Kneur:2021feo}.
In addition, the figure  suggests that the NLO RGOPT produces EoSs that converge slower to the Fermi-Dirac limit (free massless quarks) than its NNLO pQCD counterpart. This happens because the 
former procedure produces a non-trivial medium-dressed mass $\overline\eta(g,\mu)$, so that the effective masses entering the quark propagators are $m_s + \overline{\eta}(g,\mu)$ (strange sector) and $\overline{\eta}(g,\mu)$ (non-strange sector). Accordingly, at $N_f=2$ the RGOPT quark propagator already involves a non-trivial effective quark mass while, at $N_f=2+1$, the effect of including the $s$-quark physical (current) mass is more diluted. A similar behavior was initially noticed to occur within NLO pQCD  \cite{Fraga:2004gz}, where the introduction of the strange quark mass  was observed to push down the pressure at low-$\mu_B$ values ($\mu_B\lesssim 3$ GeV), softening the resulting EoS. However, later the authors of Ref. \cite{Kurkela:2009gj} found that the softening effect is less noticeable when finite masses are considered  at NNLO.  

\subsection{Quark stars\label{qs}}
\begin{figure}
\begin{subfigure}
    \centering
     \includegraphics[width=.45\textwidth]{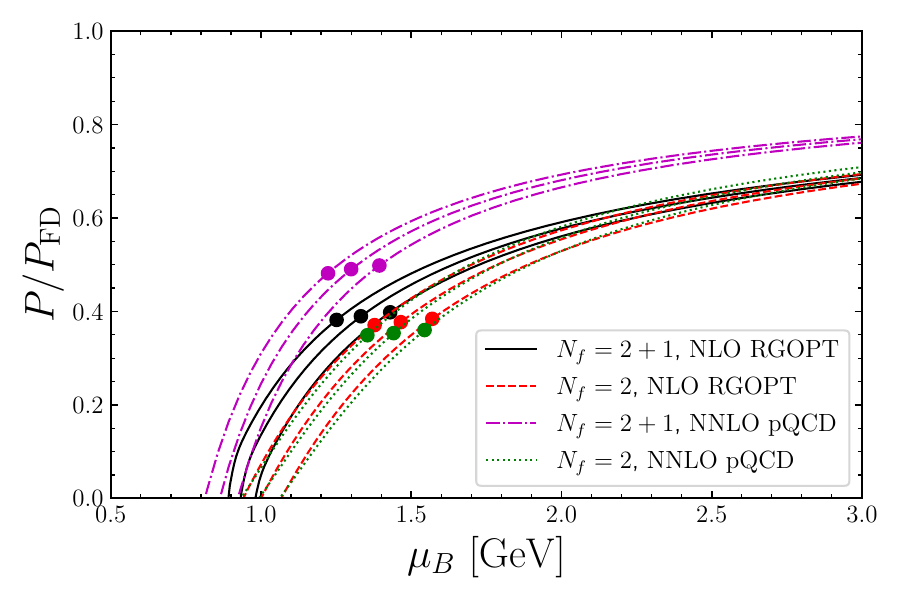} 
\end{subfigure}
\begin{subfigure}
 \centering
 \includegraphics[width=.45\textwidth]{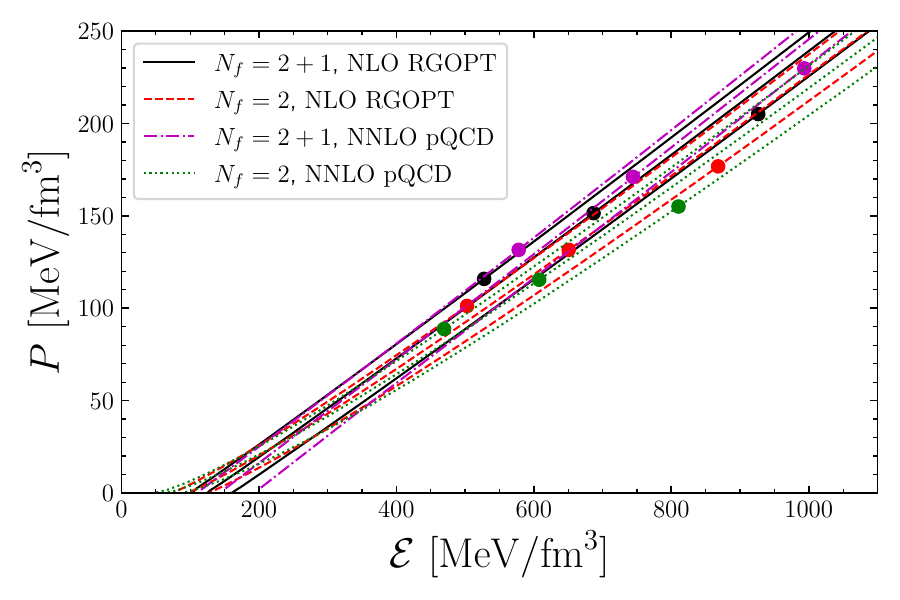} 
 \end{subfigure}
 \begin{subfigure}
     \centering
     \includegraphics[width=.45\textwidth]{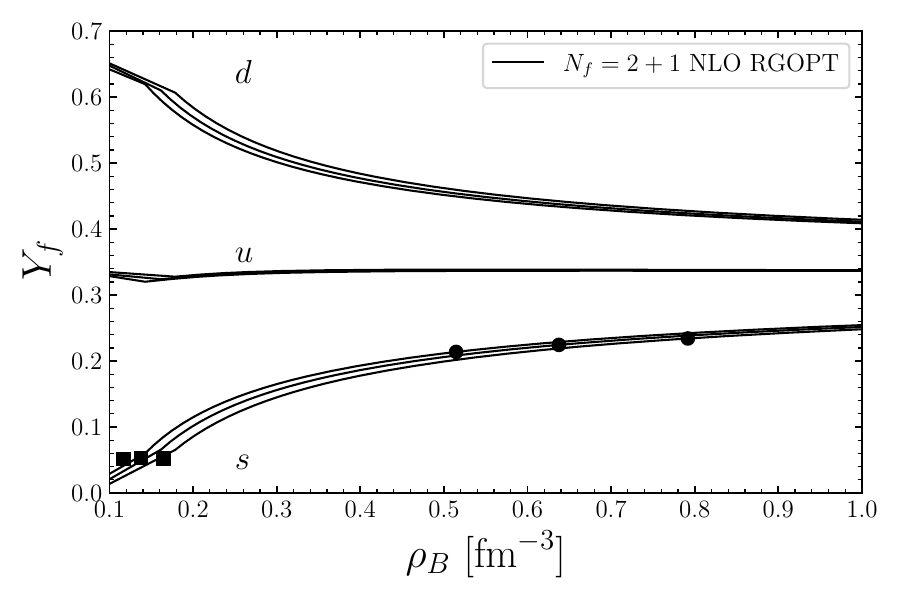}
 \end{subfigure}
\caption{Pressure versus baryon chemical potential (top left) and energy density (top right) of $\beta$-equilibrated matter for the $N_f=2+1$ (continuous lines) and $N_f=2$ (dashed lines) NLO RGOPT, and for $N_f=2+1$ (dot-dashed lines) and $N_f=2$ (dotted lines) NNLO pQCD. Bottom panel: the quark fractions  as a function of the baryonic density for the same EoS. The curves were obtained with the scale values $X$  that reproduce $M_{\rm max}=2,\, 2.3$ and $2.6 M_\odot$, given in Table \ref{tab:tabla4.1}. The dots identify the values at the center of the maximum mass star configuration. The squares in the bottom panel identify the baryonic densities at which $P=0$. }
\label{EoS_2L_Aopt_band}
\end{figure}

\begin{figure}
 \centering
 \includegraphics[width=.75\textwidth]{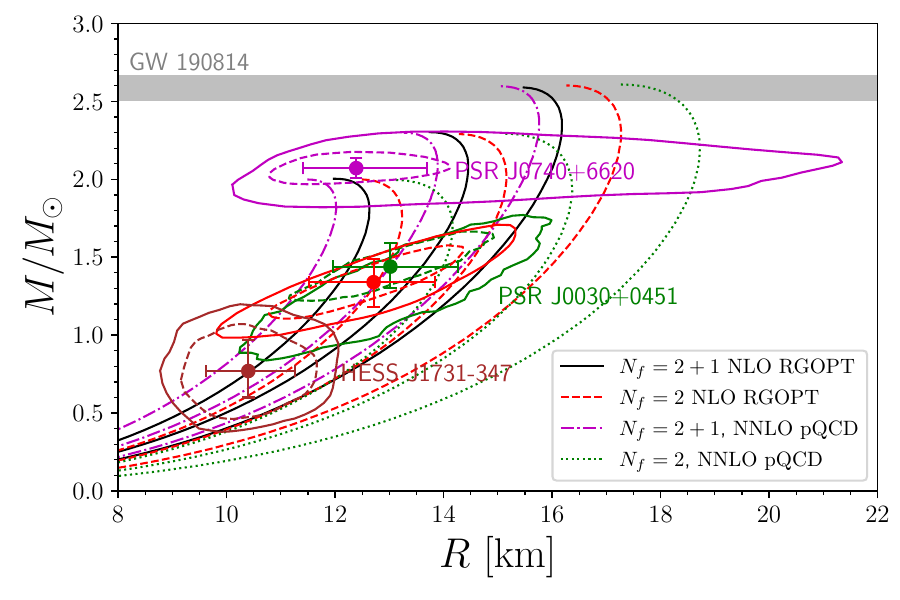} 
\caption{Mass-radius relation given by the different approximations. 
Lines reproducing maximum QS masses of $2.6 M_\odot$ correspond to the
higher values of $X$ of table \ref{tab:tabla4.1}, while the ones for $2 M_\odot$
correspond to the lowest $X$ values.
Also included are the NICER data for  pulsars PSR J0030+0451 \cite{Riley_2019,Miller:2019cac} and PSR J0740+6620 \cite{Riley:2021pdl,Miller:2021qha}, as well as  the low mass compact star HESS J1731-347 \cite{2022NatAs.tmp..224D}. In particular, the ellipses represent the 68\% (dashed) and  95\% (full) confidence interval of the 2-D  distribution in the mass-radii domain  while the error bars give the 1-D marginalized posterior distribution for the same data. A band identifying the mass of the low mass compact object associated with GW190814 \cite{LIGOScientific:2020zkf} has also been included.}
\label{Mass_Radius_Aopt}
\end{figure}

\begin{table}
\caption{QS properties predicted by  NLO $\rm RGOPT \ $ with  $N_f=2+1$ and $N_f=2$,  and by  NNLO pQCD with $N_f=2+1$ and $N_f=2$ with the $X$ scale values that reproduce  $M_{\rm max}=2,\, 2.3$ and $2.6 M_\odot$. The considered properties  are: maximum mass $M_{\rm max}$ and corresponding radius $R_{\rm max}$,  radius  of the 1.4$M_\odot$ and 0.77$M_\odot$ stars, $R_{1.4}$ and $R_{0.77}$, central baryon density $\rho_B^{c,\rm max}$ and central baryon chemical potential $\mu_B^c$, 
as well as surface  baryonic density and chemical potential  of the maximum mass configuration.}
\begin{center}
\begin{tabular}{c  |c  c  c  c c c c c c} 
       \hline
    &  $X$ & $M_{\rm max}$ &  $R_{\rm max}$  & $R_{1.4}$ & $R_{0.77}$  & $\rho_B^{c,\rm max}$  & $\mu_B^{c,\rm max}$   & $\rho_B^{\rm surf, max}$ & $\mu_B^{\rm surf, max}$\\
    &  & ($M_\odot$) &  (km) & (km) & (km) & ($\rho_0 $)  & (GeV)   & ($\rho_0 $)  & GeV\\
	\hline
	\multirow{2}{4.5em}{$N_f=2+1$ $\rm RGOPT \ $} & $3.63$ & $2.00 $ &  $12.0$ & $12.2$ & $10.4$& $4.95 $ & $1.430 $ & $1.03 $ &$0.969$\\ 
  &$3.97$& $2.30 $ & $ 13.7$ &$13.5$ & $11.4$& $3.98 $ & $1.337$ & $0.86 $ & $0.930$   \\
     &$4.30 $& $2.59 $ &$15.5$ & $ 14.6$ &$12.7$ & $3.21 $ & $1.251$ & $0.73 $ & $0.890$\\
     \hline
    \multirow{2}{4.5em}{$N_f=2+1$ $\rm pQCD \ $} & $2.95$ & $2.00 $ &  $11.2$ & $11.6$ & $9.85$& $5.48 $ & $1.394 $ & $1.31 $ & $0.923$ \\ 
  &$3.26$& $2.30 $ & $ 13.3$ &$12.9$ & $11.0$& $4.40 $ & $1.300$ & $1.04 $ & $0.863$  \\
     &$3.56 $& $2.60 $ &$15.1$ & $ 14.2$ &$12.0$ & $3.62 $ & $1.223$ & $0.84 $ & $0.813$ \\
     \hline
       \multirow{2}{4.5em}{$N_f=2$ $\rm RGOPT$} & $3.39$ & $2.00$ & $12.4$ & $12.9$ & $11.1$ &$ 4.17  $ & $1.570$ & $0.74 $ & $1.067$ \\
          &$3.80$ & $2.29 $ & $14.3$& $14.3$ & $12.2$ & $3.34   $ & $1.465$ & $0.61 $ & $1.000$ \\
     &$4.18$ & $2.60 $ & $16.3$ & $15.9$ & $13.4$ & $2.74   $ & $1.378$ & $0.49 $ & $0.941$\\
	\hline
  \multirow{2}{4.5em}{$N_f=2$ $\rm pQCD$} & $2.17$ & $2.00$ & $13.1$ & $13.9$ & $12.2$ &$ 3.92  $ & $1.544$  & $0.50 $ & $1.053$ \\
          &$2.35$ & $2.29 $ & $15.1$& $15.4$ & $13.6$ & $3.14   $ & $1.441$ & $0.39 $ & $0.982$ \\
     &$2.53$ & $2.61 $ & $17.2$ & $17.5$ & $15.1$ & $2.58   $ & $1.354$ & $0.31 $ & $0.920$ \\
	\hline
\end{tabular}\label{tab:tabla4.1}
\end{center}
\end{table}

Having discussed the EOS of $\beta$-equilibrium quark matter, for both SQM
and NSQM, we next  consider NS observations to constrain the scale $X$. We will
apply our EOS to the description of quark stars (QS). We will not consider
hybrid stars with a core of quark matter in order to avoid an extra uncertainty
through the hadronic EoS.

In Fig. \ref{EoS_2L_Aopt_band} we display the EoSs, pressure versus chemical potential and pressure versus the energy density  (top panels) and the respective quark fractions (bottom panel) for the values of $X$ that reproduce $M_{\rm max}=2,\, 2.3$ and $2.6 M_\odot$. 
The corresponding values \footnote{Notice   that the $X$ values of the RGOPT and pQCD for NSQM ($N_f=2$) slightly differ from those of Ref. \cite{Restrepo:2022wqn} since in that work the scale was chosen to be $\Lambda=X(\mu_u+\mu_d)/2$.} for both approximations in the case of SQS  and NSQS are given in Table \ref{tab:tabla4.1}. In this table, one can see the NLO RGOPT may require $X$ values  slightly higher than those of NNLO pQCD, namely $X=3.63-4.30$ for SQS and $X=3.39-4.18$ for NSQS, implying that such  stars may be composed by weakly interacting quarks ( smaller $\alpha_s(X)$), even at low $P$-values. 
In contrast, the NNLO pQCD produces the required maximum masses for values of $X$ within the canonical  range, namely $X=2.95-3.56$ for SQS and $X=2.17-2.53$ for NSQS. 
We also checked that using higher orders in the running coupling and mass
instead of Eqs.(\ref{g2L}),(\ref{msrunning}) would have very moderate, not much
visible impact on our results,
which is partly due to our use of exact two-loop running coupling, Eq.(\ref{g2L}), less sensitive
to higher order differences than the truncated power-expansion in
$\ln^{-1}(\Lambda/\mu)$ more commonly used in the literature. Moreover, note
that all our RGOPT results in Table \ref{tab:tabla4.1} lie within a range of $\mu_B$ and $X$
values such that the coupling $\alpha_s$ remains moderately perturbative, $0.35
\le \alpha_s\le 0.38$ roughly for the values of the scale at the surface of the
stars with maximum masses.

We should also comment on the configurations shown in Table
\ref{tab:tabla4.1} by taking into account the stability constraints generally
considered for quark matter: i) for stable strange matter, the  energy per
baryon at the bulk
should be below 930 MeV \cite{Bodmer:1971we,Witten:1984rs}, while,
for the same input parameters in a given model,
$ud$ quark
matter should be no more bound than iron, i.e. the  energy per baryon should be
above 930 MeV, see the discussion in Ref.\cite{Torres:2012xv}; ii) for
stable $ud$ quark matter, the energy per baryon should be below 930 MeV.
Recently, other
scenarios have been proposed within phenomenological quark models
\cite{Holdom:2017gdc,Yuan:2022dxb}, and therefore the conditions of absolute
stability are still under discussion.

Considering first the results obtained with pQCD given
in Table \ref{tab:tabla4.1}, they show stable NSQM only for $X\gtrsim 2.5$,
corresponding to $M_{\rm{max}}\gtrsim 2.56 M_\odot$. Looking at the relevant
pQCD $X$ values for SQM, the corresponding energies per baryon are below 930 MeV
in bulk for almost all values listed in Table \ref{tab:tabla4.1}, i.e. for
$X\gtrsim 3$. However, at the same $X$ values, the NSQM energies per baryon are
also below 930 MeV, but would give very large corresponding maximum masses
$M_{\rm{max}}\gtrsim 2.7 M_\odot$.
In several recent studies \cite{Annala:2021gom, Gorda:2022jvk} it is shown, using different agnostic descriptions of the neutron star EOS and imposing pQCD constraints, that the maximum mass of the NS is below $\sim 2.5 M\odot$.
Therefore, according to these studies models with $M_{\rm{max}} \ge 2.7M_\odot$
appear unfavored. However, some works propose the existence of quark stars
with masses above $2.5M\odot$: in Ref. \cite{Bombaci:2020vgw} a scenario of uds
QS with a maximum mass of $2.5 -2.6M_\odot$ is associated to the binary neutron
star merger GW190814, and in Ref. \cite{Zhang:2020jmb} the authors obtain within
an interacting quark model $M_{\rm{max}} \le 3.23M_\odot$, a value of the order
proposed by Rhoades and Ruffini \cite{Rhoades:1974fn} as maximum NS mass
determined just from the Einstein’s theory of relativity, the principle of
causality, and Le Chatelier's principle.
Moreover, note that pQCD constraints are treated differently in Ref.
\cite{Bai:2024amm}, where quark stars are discussed.

In the RGOPT case for SQM, the Bodmer-Witten hypothesis is fulfilled {and NSQM
is less bound than iron} as far as $3.97\lesssim X\lesssim 4.24$, corresponding
to $2.30 M_\odot\lesssim M_{\rm{max}}\lesssim 2.54 M_\odot$, thus in a more
constrained range than for pQCD, for which SQM  matter is stable in the entire
range of 2-2.5$M_\odot$.
The upper limit is obtained taking the same $X$ range  for NSQM: we obtain
energies per baryon equal to
$0.969 (0.925)$ GeV, respectively, for $X=3.97 (4.30)$. More precisely, one has
stable NSQM for $X\gtrsim 4.24$, corresponding to $M_{\rm{max}}\gtrsim 2.69
M_\odot$, and therefore the condition $X< 4.24$ must be imposed.

The two scenarios obtained with both approaches (pQCD or RGOPT) for
stable NSQM do not appear to be realistic according to several different
studies that propose maximum masses of the order of 2.15-2.35 $M_\odot$
\cite{Margalit:2017dij,Rezzolla:2017aly,Ruiz:2017due,Annala:2021gom}  and
generally below 2.5 $M_\odot$ \cite{Annala:2021gom}. Therefore, within the
possible limitations of our level of approximations, and considering
realistic maximum masses between 2.3 and 2.5 solar masses, our descriptions
seem to agree with the Bodmer-Witten conjecture on SQM, with $ud$ quark matter
being less bound than iron.
This could suggest that pure quark stars would be made of $uds$
matter, while $ud$ quark
stars would carry an outer layer of hadronic matter or could be quark stars.
These results indicate that the composition of dense matter is still not clear
and further studies are required.

With respect to remnant scale dependence, it is important to notice that these values indicate that the RGOPT is moderately more stable to scale variations than pQCD. Indeed, to reproduce the maximum mass variation $\Delta M_{\rm max}= (2.6 - 2.0) M_\odot$ for SQS this method requires a relative scale variation $\Delta X=0.67$, which is slightly higher than the one required by pQCD, $\Delta X=0.61$. A similar situation occurs in the case of NSQS when the OPT requires $\Delta X=0.79$ while pQCD only requires $\Delta X=0.36$.   On the different panels of  Fig. \ref{EoS_2L_Aopt_band}, the dots identify the central chemical potential or baryonic density of the maximum mass configurations. At the same time, the squares on the bottom panel indicate the $P=0$ condition that defines the QS surface. Note that on the surface $s$-quarks are already present. 

Fig. \ref{Mass_Radius_Aopt} shows the mass-radius relations obtained with the EoSs of Fig \ref{EoS_2L_Aopt_band} together with the predicted masses and radii of the pulsars PSR J0030+0451 \cite{Riley_2019,Miller:2019cac} and  PSR J0740+6620 \cite{Riley:2021pdl,Miller:2021qha} as well as the compact object HESS J1731-347 \cite{2022NatAs.tmp..224D}. For reference, a band identifying the low mass compact object associated with GW190814 \cite{LIGOScientific:2020zkf} is also shown. 
Comparing the results for NSQS with SQS, we find that the NLO RGOPT results show a slight change in the neutron star radii, which are lower for the SQS. This could be expected, since the onset of a new degree of freedom softens the EoS. As a consequence of the softening, in order to describe two solar mass stars or above, larger scale values $X$ are required. In the case of SQM the $X$ values 
are slightly larger than the ones obtained for pQCD.
However, we note that for the maximum mass range considered [2:2.6]$M_\odot$, the NLO RGOPT  predictions for both NSQS and SQS  properties that are compatible with the present NS observations. More stringent constraints are necessary to distinguish the different scenarios. Considering the choice of scale adopted, the descriptions within both pQCD and RGOPT for $N_f=2+1$ are compatible with each other, with NLO RGOPT predicting slightly larger radii.  Compared with  NNLO pQCD, NLO RGOPT  gives a smaller reduction in the radii predicted by SQS (not more than 1~km) compared to NSQS.   This is explained by the large difference between the pQCD expressions for $m\ne 0$ and $m=0$, while RGOPT produces similar expressions in both cases, since even in the massless sector, the  prescription introduces a resummed medium-dressed mass, $\overline\eta(g,\mu)$.

\section{Conclusions}

We have considered  the RGOPT resummation approach, at NLO, in order to compare strange and non-strange pure QS. The latter case was originally investigated in Ref. \cite{Restrepo:2022wqn}, when massless up and down quarks ($N_f=2$) were considered to represent the relevant degrees of freedom for the QCD EoS. 
Here, the $N_f=2$ application was extended to the $N_f=2+1$ case to include a new degree of freedom represented by the (massive) strange quark. 
Accordingly, the RG operator contains the anomalous mass dimension $\gamma_{m_s} \partial/ \partial m_s$ that was absent in previous RGOPT applications to QCD where the chiral limit was considered.
 
Following the conventional practice, we have chosen the renormalization scale to be density dependent, $\Lambda = X \mu_B/3$. 
Consequently, parameters such as the strong coupling and the strange quark mass also run with $\mu_B$ implying that the pressure needs to be modified in order to produce thermodynamic consistent predictions. 

After carrying out those modifications, in accordance with Ref. \cite{Restrepo:2022wqn}, we 
have  determined the EoS representing the two scenarios, with and without the $s$-quark. 
In more academic applications, the value of the parameter characterizing the scale, is generally chosen conventionally to be $X \in [1,4]$ with $X=2$ representing the so-called central scale. 
In the present work, this arbitrary scale parameter was fixed in a more phenomenological way by selecting the values which  give maximum QS masses of order $M_{\rm max} = 2,\,  2.3$ and $2.6\, M_\odot$ which  represent some of the limits often considered in the literature. For the NLO RGOPT EoS the required values are, respectively, $X=3.63, \, 3.97$ and  $\, 4.30$ when $N_f=2+1$ and $X=3.39,\, 3.80$ and $4.18$ when $N_f=2$ whereas for the NNLO pQCD EoS the values are $X=2.95,\, 3.26$ and $3.56$ when $N_f=2+1$ and $X=2.17,\, 2.35$ and $2.53$ when $N_f=2$.
Accordingly, within the RGOPT framework, massive QS can only be produced at relatively larger scales (smaller couplings) compared with pQCD. The introduction of an extra degree of freedom, the $s$-quark, softens the EoS and to attain the same maximum masses the $N_f=2+1$ scenario requires larger $X$ scales in both descriptions. 
With these choice of the scale, it was shown that the EoS obtained are in accordance with properties of several recent neutron star mass-radius data, in particular, with data for  pulsars PSR J0030+0451 \cite{Riley_2019,Miller:2019cac} and PSR J0740+6620 \cite{Riley:2021pdl,Miller:2021qha}, and for the low mass compact star HESS J1731-347. 
To understand these results, let us first recall that the introduction of finite quark masses tends to reduce the pressure predicted by both methods, as discussed earlier. This effect is amplified in the RGOPT approach, since the effective quark masses are $m_s+\overline\eta$ (in the strange heavy sector) and $\overline\eta$ (in the non-strange light sector), implying that the corresponding EoS for the same scale $X$ and baryonic chemical potential reach lower pressures than their pQCD counterparts.
Therefore, the RGOPT predicts that QS with high masses can only be formed when the scale is rather high. Comparing the $N_f=2+1$ and $N_f=2$ RGOPT mass-radii curves  one can see that, despite their similarities and good agreement with observations, the description of strange QS requires $X>3.63$ implying that the coupling takes relatively lower numerical values. The $N_f=2+1$ NNLO  pQCD  description  of SQS also calls for somewhat higher $X$ values which, however, lie within  the more canonical range. 
Note also that the RGOPT is slightly more stable to scale variations than pQCD. In particular, to reproduce the maximum mass variation $\Delta M_{\rm max}= (2.6 - 2.0) M_\odot$ corresponding to possible different scenarios for SQS, NLO RGOPT requires a relative scale variation of  $\Delta X=0.67$, about $10 \%$ larger than the variation occurring in  pQCD, $\Delta X=0.61$. 
We have also studied more precisely the stability of SQM and NSQM scenarios
upon examining the energy per baryon for our different QM approximations.
For both pQCD and RGOPT, $ud$ quark matter appear less bound than iron, except
for very large masses $M_{\rm{max}}\gtrsim 2.6 M_\odot$.
On the other hand, the NLO RGOPT predictions are in agreement with the SQM
Bodmer-Witten conjecture within the mass range $2.30 M_\odot\lesssim
M_{\rm{max}}\lesssim 2.54 M_\odot$. At the same time, they satisfy the condition
that $ud$ matter is less bound than iron.
Note,  that it is  difficult to assert firm
conclusions about Bodmer-Witten conjecture without a truly non-perturbative contribution
to the QCD vacuum energy, that neither pQCD nor RGOPT incorporate.

At this point, we must emphasize that a potentially
more reliable RGOPT prediction will require a computation extended at NNLO, since this next perturbative order contains sizeable massive contributions \cite{Kurkela:2009gj}
from three-loop graphs with three-gluon vertices and correction to the quark-gluon vertex, 
which may turn out to  affect the 
softness of the EoS with respect to the NLO case. While 
the RGOPT evaluation of the basic NNLO pressure with massive quarks 
was recently obtained \cite{Fernandez:2024ilg}, extending those results
to consistently account for chemical equilibrium and thermodynamic consistency is somewhat involved and beyond the scope of the present work. Finally, we have observed that the method, once generalized for the inclusion of a genuine physical (current) mass $m_s(\Lambda)$, produces results which are slightly less sensitive to scale variations than those provided by pQCD, just like what happens when massless theories are considered 
\cite{Kneur:2019tao, Kneur:2021dfo,Kneur:2021feo, Restrepo:2022wqn}. 

\section*{Acknowledgments}
{T.E.R acknowledges support from Funda\c c\~ao Carlos Chagas Filho de Amparo \` a Pesquisa do Estado do Rio de Janeiro (FAPERJ), Process SEI-260003/019683/2022. M.B.P. is partially supported by Conselho
Nacional de Desenvolvimento Cient\'{\i}fico e Tecnol\'{o}gico (CNPq),
Process  No.  307261/2021-2  and 403016/2024-0.  T.E.R and M.B.P are also partially supported by
Instituto  Nacional  de  Ci\^encia  e Tecnologia de F\'{\i}sica
Nuclear e Aplica\c c\~{o}es  (INCT-FNA), Process No.  464898/2014-5. C.P. received support from Fundação para a Ciência e a Tecnologia (FCT), I.P., Portugal, under the  projects UIDB/04564/2020 (doi:10.54499/UIDB/04564/2020), UIDP/04564/2020 (doi:10.54499/UIDP/04564/2020), and 2022.06460.PTDC (doi:10.54499/2022.06460.PTDC). }

\bibliography{bibliography.bib}
\end{document}